\documentclass[aps,prl,twocolumn]{revtex4}
\usepackage{amsmath,graphics,graphicx,float}
\begin{document}

\title{Nonequilibrium thermal entanglement}

\author{Luis Quiroga, Ferney J. Rodr\'{\i}guez}
\affiliation{Departamento de F\'{\i}sica, Universidad de Los Andes, A.A.4976,
Bogot\'a D.C., Colombia}
\author{Mar\'{\i}a E. Ram\'{\i}rez, Roberto Par\'{\i}s}
\affiliation{Departamento de F\'{\i}sica, Universidad Javeriana,
Bogot\'a D.C., Colombia}

\date{\today}

\begin{abstract}
Results on heat
current, entropy production rate and entanglement are reported for
a quantum system coupled to two different
temperature heat reservoirs.
By applying a temperature gradient, different quantum
states can be found with exactly the same amount of entanglement
but different purity degrees and heat currents. Furthermore, a
nonequilibrium enhancement-suppression transition behavior of the entanglement
is identified.
\end{abstract}
\pacs{PACS numbers: 03.67.Mn,03.65.Ud,65.40.Gr}
\maketitle

\section{I. Introduction}
Quantum thermodynamics \cite{rego,mahler} is starting to throw
light on universal behaviors of nanosystems. Specifically, new
possibilities arising in nonequilibrium situations, with dominant quantum
coherences, are emerging \cite{scully}. Currently,
non-local quantum correlations (entanglement) are being considered
in a vast variety of scenarios as they are key ingredients for
novel and non-conventional forms of communication, information
processing and computation. For potential large-scale
applications, where condensed matter systems are of prime
importance, thermal interactions with specific environments are
unavoidable. Thus a clear connection between quantum information
aspects and thermal magnitudes has to be elucidated. Furthermore,
the biological frontier of physics imposes to address the question
of quantum features survival in noisy as well as in nonequilibrium
conditions \cite{phillips}.

For quantum systems in contact with heat reservoirs at unique
and fixed temperature, the equilibrium thermal entanglement has
been extensively studied \cite{vedral,wang,kamta,canosa}. Until
now, most of the emphasis in the study of thermal entanglement has
been confined to equilibrium situations. Entanglement in
nonequilibrium quantum systems has been scarcely considered and
thus a proper description of thermal entanglement in the presence
of matter and/or energy currents is still lacking. Recently,
Eisler et al. \cite{eisler} has calculated the Von Neumann entropy
of a block of spins in a XX spin chain in the presence of an
energy current, showing that an enhancement of the amount of
entanglement due to an energy current is possible. The energy
current is modelled by adding an extra term to the spin chain
Hamiltonian for simulating a steady-state current in a
thermodynamic closed system. However, the issue of entanglement
behavior in true nonequilibrium conditions of a thermodynamic open
system remains untouched.

Two
coupled qubits in thermal contact with different heat baths is a
system not only of theoretical interest but a common place in
nanophysics. In semiconductor quantum dots the transfer of quantum
information between nuclear spins and electronic spins has been
recently considered \cite{quiroga,imamoglu1,loss}. The nuclear
spin is generally weakly coupled to its environment while the
electronic spin is strongly coupled to a great variety of degrees
of freedom within the solid. In this way, the effective
environments are different for both kind of spins. Besides that,
nuclear magnetic resonance techniques allow the cooling of nuclear
spins in a controlled manner \cite{imamoglu2,lai} without
significatively affecting the electronic spins, thus creating two
reservoirs at effective different temperatures. On the other hand,
superconductor qubits can be easily designed to be coupled to
different environments. For instance, inductively coupled
superconductor flux qubits in contact with two different
environments has been recently analyzed in Ref.\cite{storcz}.
The aim of the present paper is to correlate thermodynamical
nonequilibrium steady-state features with entanglement properties
of quantum nanosystems. In doing so, we shall consider a
quantum system in a nonequilibrium condition
for which the amount of entanglement can be exactly evaluated: two
interacting qubits (spins) in contact with two heat reservoirs at
different temperatures. In this case, entanglement can be evaluated
for any mixed state by using the concurrence \cite{wooters}.
Indeed, the model system to be considered in the present
work should be useful for a large variety of physical set ups
aiming to explore the relationship between quantum informational
entropy and thermodynamic entropy at the atomic scale. Whether one
can reveal universal features in irreversible processes of open
quantum systems is of great significance.

\section{II. Formalism}
The central quantum nanosystem is described by a
Hamiltonian $\hat{Q}$, interacting with two heat reservoirs
which are assumed to be in a permanent
thermodynamical equilibrium at $\beta_i=1/k_BT_i, i=1,2$
($k_B=\hbar=1$) with internal Hamiltonians $\hat{R}_i$.
 The total Hamiltonian is then
\begin{eqnarray}
\hat{H}=\hat{Q}+\hat{R}_1+\hat{R}_2+\hat{S}_1+\hat{S}_2
\end{eqnarray}
where the nanosystem is simultaneously coupled with both
reservoirs through terms $\hat{S}_1$ y $\hat{S}_2$. The nanosystem
+ reservoirs is described by a density operator
satisfying the Liouville equation
$\frac{d\hat{\gamma}}{dt}=-i[\hat{H},\hat{\gamma}]$. We assume
that the coupling strengths of the central quantum system to the
reservoirs are weak so that the full density operator
$\hat{\gamma}$ can be expressed as $\hat{\gamma}(t)=\hat{\rho}(t)
\hat{\rho}_1 \hat{\rho}_2$ where each reservoir is described by
its own canonical equilibrium density operator
$\hat{\rho}_i=e^{-\beta_i \hat{R}_i}/Tr_{R_i}\{e^{-\beta_i
\hat{R}_i} \} \}$ and $\hat{\rho}(t)$ is the reduced density
operator for the quantum system of interest. The couplings
nanosystem-reservoirs are written as
\begin{eqnarray}
\hat{S}_1+\hat{S}_2=\sum_{j=1}^2\sum_{\mu}
\hat{V}_{j,\mu}\hat{f}_{j,\mu}=\sum_{j=1}^2\sum_{\mu}
\hat{V}^{\dag}_{j,\mu}\hat{f}^{\dag}_{j,\mu}
\end{eqnarray}
where the nanosystem operators $\hat{V}_{j,\mu}$ are taken to
satisfy
$[\hat{Q},\hat{V}_{j,\mu}]=\omega_{j,\mu}\hat{V}_{j,\mu}$
and the operators $\hat{f}_{j,\mu}$ act on the reservoir degrees
of freedom ($j=1,2$). Within the framework of the Born-Markov
approximation \cite{goldman}, the equation of motion for
$\hat{\rho}(t)$ is given by
\begin{eqnarray}
\nonumber
\frac{d\hat{\rho}}{dt}=-i[\hat{Q},\hat{\rho}(t)]
-\sum_{j=1}^2\sum_{\mu,\nu}J_{\mu,\nu}^{(j)}(\omega_{j,\nu})\\
\{[\hat{V}_{j,\mu},[\hat{V}^{\dag}_{j,\nu},\hat{\rho}]]
-(1-e^{\beta_j\omega_{j,\nu}})[\hat{V}_{j,\mu},\hat{V}^{\dag}_{j,\nu}\hat{\rho}]\}\label{Eq:r1}
\end{eqnarray}
where the spectral density of the $j$-th reservoir is
\begin{eqnarray}
J_{\mu,\nu}^{(j)}(\omega_{j,\nu})=\int_0^{\infty} d\tau
e^{i\omega_{j,\nu}\tau}Tr_{R_j}\{\hat{\rho}_j\bar{f}^{\dag}_{j,\nu}(\tau)\hat{f}_{j,\mu}\}
\end{eqnarray}
with
$\bar{f}^{\dag}_{k,\nu}(\tau)=e^{-i\hat{R}_k\tau}\hat{f}^{\dag}_{k,\nu}e^{i\hat{R}_k\tau}$.

We will here be concerned with the simplest possible scenario
where clear relations between informational and thermodynamic
entropies could be found. To set up our model system in a general
context, we consider a nanosystem composed of two interacting
qubits as described by the Hamiltonian
\begin{eqnarray}
\hat{Q}=\sum_{\alpha=1}^2\frac{\epsilon_{\alpha}}{2}\hat{\sigma}_{\alpha,z}+
K(\hat{\sigma}_{1}^+\hat{\sigma}_{2}^-+\hat{\sigma}_{1}^-\hat{\sigma}_{2}^+)+
K^{\prime}\hat{\sigma}_{1,z}\hat{\sigma}_{2,z}\label{Eq:q}
\end{eqnarray}
where $\hat{\sigma}_{\alpha,z}$ and $\hat{\sigma}_{\alpha}^{\pm}$
denote Pauli matrices. The inter-qubit coupling is ferromagnetic
when $K,K^{\prime}<0$ and antiferromagnetic when $K,K^{\prime}>0$.
This type of Hamiltonian encompasses three well-known spin models:
it turns into the isotropic Heisenberg-like coupling for
$K=K^{\prime}$, the isotropic XX-like model for $K^{\prime}=0$ and
the Ising-like model for $K=0$. The eigenenergies and eigenstates
corresponding to Eq.(\ref{Eq:q}) are: $|s_1\rangle=|0,0\rangle$
($E_1=-\frac{\epsilon_1+\epsilon_2}{2}+K^{\prime}$),
$|s_2\rangle=|1,1\rangle$
($E_2=\frac{\epsilon_1+\epsilon_2}{2}+K^{\prime}$),
$|s_3\rangle={\rm cos}(\theta/2)|1,0\rangle+{\rm
sin}(\theta/2)|0,1\rangle$ ($E_3=\alpha-K^{\prime}$) and
$|s_4\rangle=-{\rm sin}(\theta/2)|1,0\rangle+{\rm
cos}(\theta/2)|0,1\rangle$ ($E_4=-\alpha-K^{\prime}$), with
$\alpha=\sqrt{K^2+\frac{(\epsilon_1-\epsilon_2)^2}{4}}$ and ${\rm
tan}\theta=2K/(\epsilon_1-\epsilon_2)$. We consider each qubit in
contact with its own boson heat reservoir through a term of the
form
\begin{eqnarray}
\hat{S}_j=\hat{\sigma}_j^+\sum_n
g^{(j)}_n\hat{a}_{n,j}+\hat{\sigma}_j^-\sum_n
g^{(j)*}_n\hat{a}^{\dag}_{n,j}\quad ,\quad j=1,2
\end{eqnarray}
where $\hat{a}^{\dag}_{n,j}$ creates an excitation in mode $n$ of
reservoir $j$ with a coupling strength $g^{(j)}_n$.

The {\it nonequilibrium} steady-state density matrix (designed
simply as $\hat{\rho}$ from now on), must satisfy
$\frac{d\hat{\rho}}{dt}=-i[\hat{Q},\hat{\rho}]=0$ in
Eq.(\ref{Eq:r1}), which yields to
${\cal L}_1(\hat{\rho})+{\cal L}_2(\hat{\rho})=0$
where the Lindblad or relaxation super-operators are given by
\begin{widetext}
\begin{eqnarray}
\nonumber {\cal
L}_j(\hat{\rho})=&-&\sum_{\mu=1}^4J^{(j)}(\omega_{\mu})
\{-\hat{V}_{j,\mu}\hat{\rho}\hat{V}^{\dag}_{j,\mu}+\hat{\rho}\hat{V}^{\dag}_{j,\mu}\hat{V}_{j,\mu}+
e^{\beta_j\omega_{\mu}}(\hat{V}_{j,\mu}\hat{V}^{\dag}_{j,\mu}\hat{\rho}-\hat{V}^{\dag}_{j,\mu}\hat{\rho}\hat{V}_{j,\mu})\}\\
&-&\sum_{\mu=1}^4J^{(j)}(-\omega_{\mu})
\{-\hat{V}^{\dag}_{j,\mu}\hat{\rho}\hat{V}_{j,\mu}+\hat{\rho}\hat{V}_{j,\mu}\hat{V}^{\dag}_{j,\mu}+
e^{-\beta_j\omega_{\mu}}(\hat{V}^{\dag}_{j,\mu}\hat{V}_{j,\mu}\hat{\rho}-
\hat{V}_{j,\mu}\hat{\rho}\hat{V}^{\dag}_{j,\mu})\}\label{Eq:ss1}
\end{eqnarray}
\end{widetext} for $j=1,2$. In the latter expression
$\omega_1=E_2-E_3, \hat{V}_{j,1}=(\delta_{j,2}{\rm
cos}(\theta/2)+\delta_{j,1}{\rm sin}(\theta/2))|s_2 \rangle\langle
s_3|$; $\omega_2=E_2-E_4, \hat{V}_{j,2}=(-\delta_{j,2}{\rm
sin}(\theta/2)+\delta_{j,1}{\rm cos}(\theta/2))|s_2 \rangle\langle
s_4|$; $\omega_3=E_3-E_1, \hat{V}_{j,3}=(\delta_{j,1}{\rm
cos}(\theta/2)+\delta_{j,2}{\rm sin}(\theta/2))|s_3 \rangle\langle
s_1|$; $\omega_4=E_4-E_1, \hat{V}_{j,4}=(-\delta_{j,1}{\rm
sin}(\theta/2)+\delta_{j,2}{\rm cos}(\theta/2))|s_4 \rangle\langle
s_1|$ and
$J^{(j)}(-\omega_{\mu})=e^{\beta_j\omega_{\mu}}J^{(j)}(\omega_{\mu})$.
Two limiting cases can be easily analyzed: (i) No inter-qubit
coupling, $K=K^{\prime}=0$ ($\theta=0$) which leads to
$\omega_1=\omega_4=\epsilon_2$ and $\omega_2=\omega_3=\epsilon_1$.
Each qubit reaches a local equilibrium with its own heat reservoir
yielding to a direct product form of the density matrix and thus
no-entanglement. (ii) Coupled qubits, $K,K^{\prime}\neq 0$, in
contact with two independent reservoirs at identical temperatures,
$\beta_1=\beta_2=\beta$. A reduced density matrix results which
has the thermodynamical canonical form for a system described by
internal Hamiltonian $\hat{Q}$ at equilibrium with a thermal bath
at inverse temperature $\beta$, as it should be.

\section{III. Results and Discussion}
Consistently with the Born-Markov approximation, we adopt a
Weisskopf-Wigner-like expression such as
$J^{(j)}(\omega_{\mu})=\Gamma_j(\omega_{\mu}) n_j(\omega_{\mu})$
where $\Gamma_j(\omega_{\mu})$ depends on both the
nanosystem-$j$th-reservoir coupling strength and the reservoir
internal structure. On the other hand,
$n_j(\omega_{\mu})=(e^{\beta_j\omega_{\mu}}-1)^{-1}$ denotes the
thermal mean value of the number of excitations in reservoir $j$
at frequency $\omega_{\mu}$. For the sake of simplicity, we take
identical and frequency independent couplings, thus
$\Gamma_1(\omega)=\Gamma_2(\omega)=\Gamma$.

From Eq.(\ref{Eq:ss1}) the {\it
nonequilibrium} steady-state density matrix is obtained as given
by the diagonal matrix $\hat{\rho}=diag\{ \rho_{1,1}, \rho_{2,2},
\rho_{3,3}, \rho_{4,4} \}$ in the basis of eigenstates of
$\hat{Q}$. Although, it can be analytically expressed we will not
go here into the details as its explicit form is cumbersome
\cite{qrrp}. Instead, we shall analyze some important special
situations.

\subsection{A. Symmetric case,
$\epsilon_1=\epsilon_2=\epsilon$}
In this case $\hat{\rho}$ can be
written in terms of a simple universal function
$e(\omega)=\frac{n_1(\omega)+n_2(\omega)}{1+n_1(\omega)+n_2(\omega)}\leq
1$ (energies and temperatures in units of interqubit coupling
$K=1$). In the strong coupling case ($\epsilon < 1$) we found
\begin{eqnarray}
\nonumber
\rho_{1,1}=\frac{e_1}{2}\left ( 1-\frac{e_2}{2} \right )&;&
\rho_{2,2}=\left ( 1-\frac{e_1}{2} \right )\frac{e_2}{2}  \\
\rho_{3,3}=\frac{e_1}{2}\frac{e_2}{2}&;&
\rho_{4,4}=\left ( 1-\frac{e_1}{2} \right )\left (
1-\frac{e_2}{2} \right ) \label{Eq:rr0}
\end{eqnarray}
where $e_j=e(\omega_j)$ with $\omega_1=\omega_4=|\epsilon-1|$ and
$\omega_2=\omega_3=\epsilon+1$. In the weak coupling case
($\epsilon
> 1$) the following interchanges have to be made:
$\rho_{1,1}\leftrightarrow \rho_{4,4}$ and
$\rho_{2,2}\leftrightarrow \rho_{3,3}$. Thus, the nonequilibrium
concurrence is $C=2Max\{0, |\rho_{3,3}-\rho_{4,4}|/2
-\sqrt{\rho_{1,1}\rho_{2,2}}\}$.

Let us first discuss the {\it equilibrium} ($T_1=T_2=T$) thermal
entanglement behavior for a system governed by Hamiltonian
(\ref{Eq:q}) \cite{vedral,wang,kamta,canosa}. An analytical
expression can be found for the equilibrium concurrence as
$C_{eq}(T)=\frac{{\rm sinh}(1/T)-1}{2{\rm cosh}(\omega_1/2T){\rm
cosh}(\omega_2/2T)}$. This last expression is interesting because
it implies an universal form (independent of $\epsilon$) for the
sudden death of the equilibrium concurrence at the temperature
$T_c=1.1346$. For $\epsilon < 1$, the two-qubit concurrence
decreases from 1 to 0 as the temperature increases up to $T_c$;
for $\epsilon
> 1$, the concurrence increases from 0 to some maximum before vanishing
at $T_c$. It is also known that the concurrence decreases
monotonically as the qubit splitting increases for any temperature
and vanishes exponentially with increasing $\epsilon$. All these
features are independent of the specific nature of the reservoirs.

The behavior of quantum states, for any qubit internal splitting
and reservoir temperatures, can be displayed in a single plot
$e_1$-$e_2$, as it is shown in Fig. 1. Although, the quantum state
variation is described by the same curve for $\epsilon$ and
$1/\epsilon$, the border lines separating the entangled from the
unentangled regions are different: the green curve corresponds to
$\epsilon < 1$ while the red curve corresponds to $\epsilon > 1$.
The black line represents equilibrium quantum states for both
$\epsilon=1/3$ as well as for $\epsilon=3$. The shifting of the
quantum state with the temperature gradient, $\Delta T=T_1-T_2$,
is depicted by the circles, for the same average temperature,
$T_M=(T_1+T_2)/2=1$. It is evident that the temperature gradient
shifts the state from the entangled zone to the unentangled zone.
However, this behavior can be reversed at low temperatures for
$\epsilon > 1$ as it is to be discussed below.

\begin{figure}
\includegraphics[width=0.7\columnwidth]{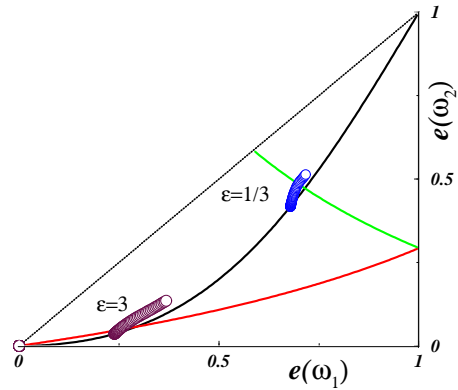}
\caption{ (Color online) Quantum state parameter space ($e_1$,
$e_2$). The black line denotes equilibrium states. Green line:
entangled-unentangled border for $\epsilon < 1$. Red line:
entangled-unentangled border for $\epsilon > 1$. Circles represent
the state shift under a temperature gradient from the same mean
temperature $T_M=1$. Blue circles: $\epsilon = 1/3$, brown
circles: $\epsilon = 3$. }
\end{figure}

In the {\it linear nonequilibrium limit} (LNEL), $\Delta T \ll 1$,
the concurrence can be written as $C(\Delta T)=Max\{0,
C_{eq}(T_M)-\alpha \Delta T^2\}$ where the coefficient $\alpha$ is
a function of the average temperature as well as the qubit
internal splitting. The equilibrium concurrence is displayed in
the insets of Figs. 2-a and 2-b. In the strong coupling limit
($\epsilon < 1$) $\alpha
> 0$ thus the concurrence is always a decreasing function of the
temperature gradient $\Delta T$. By contrast, in the weak coupling
limit ($\epsilon
> 1$) there is a transition mean temperature for which $\alpha$ changes the
sign. Thus, a low temperature region can be found where $\alpha <
0$ for which a gradient temperature produces an increasing of the
concurrence as compared with the equilibrium case. The degree of
mixing of the quantum state can be characterized by the linear
entropy as defined by $S_L=(4/3)(1-Tr\{\hat{\rho}^2\})$. The low $\Delta
T$ limit of the linear entropy can also be expanded as $S_L(\Delta
T)=S_{L,eq}(T_M)+\alpha' \Delta T^2+O(\Delta T^4)$. The
coefficient $\alpha'>0$, for both interqubit coupling cases, is
also illustrated in Figs. 2-a and 2-b. Note that while a
temperature gradient can produce, in a limited temperature
interval, an enhancement of the concurrence it always yields to a
more mixed state. This result will permit to prepare a great
variety of quantum states with practically any combination of
entanglement and purity degree by varying the temperature of only
one heat reservoir.
\begin{figure}
\includegraphics[width=0.7\columnwidth]{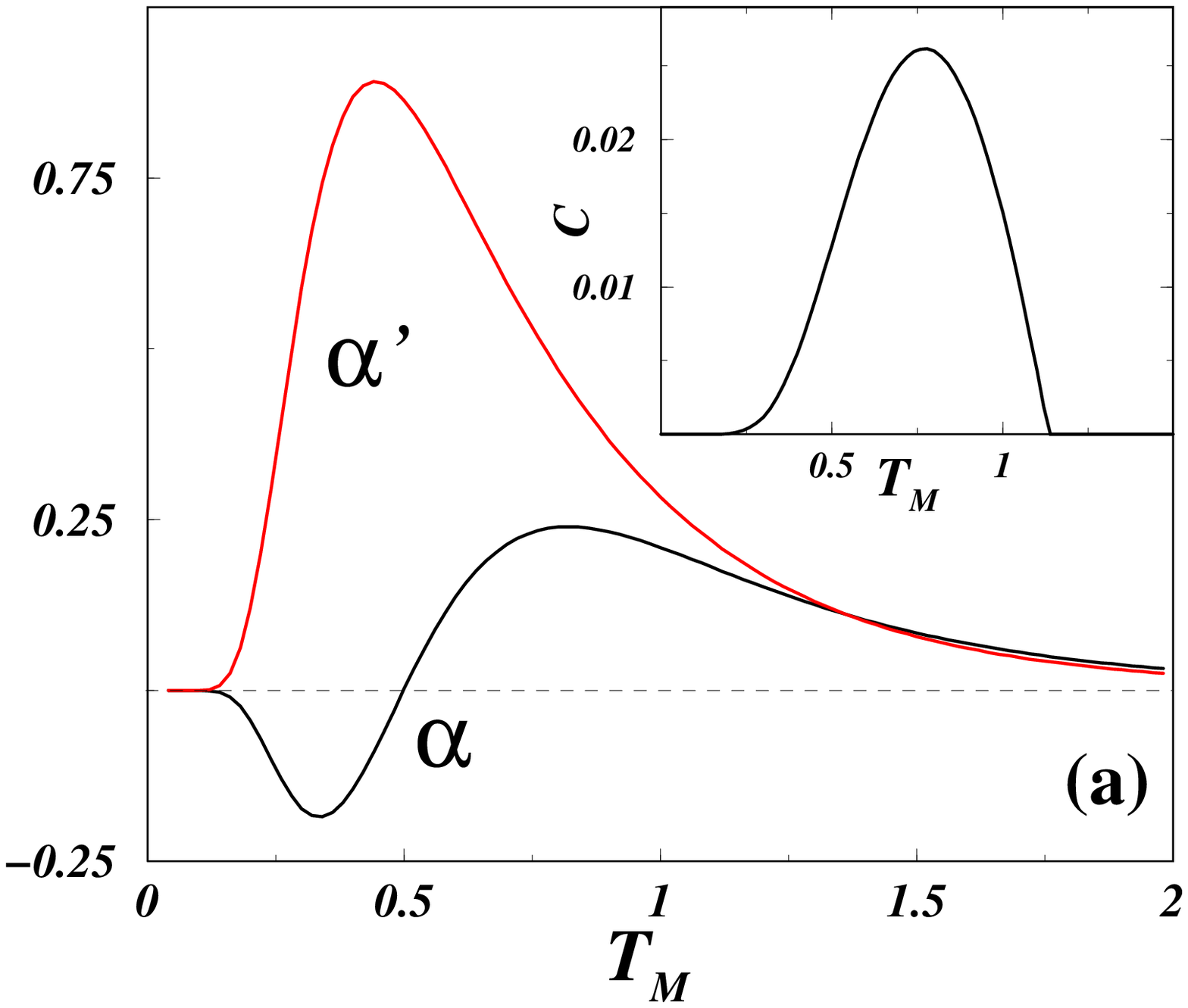}
\includegraphics[width=0.7\columnwidth]{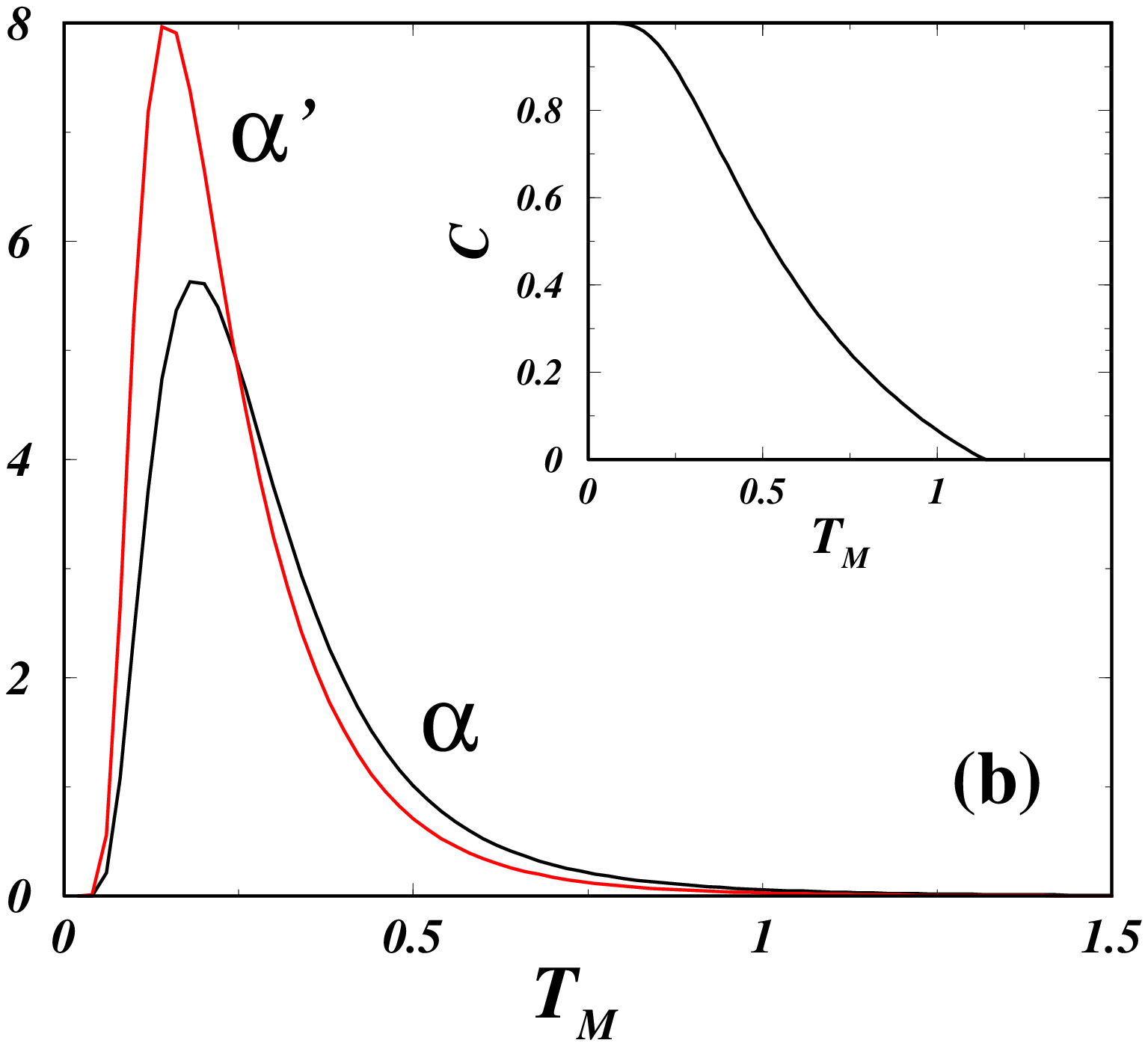}
\caption{ (Color online) LNEL coefficients $\alpha$ and $\alpha'$,
denoting the second order in $\Delta T$ variation of the
concurrence and linear entropy, respectively: (a) $\epsilon = 3$,
(b) $\epsilon = 1/3$. Insets: Equilibrium concurrence.}
\end{figure}

The relationship between a nonequilibrium thermodynamical quantity
such as the heat current and a central quantum information concept
such as entanglement is now addressed. We start by calculating the
heat current as ${\cal J}_j=Tr\{ \hat{Q} {\cal L}_j(\hat{\rho})\}$
\cite{breuer}, which in the symmetric case yields to
\begin{eqnarray}
{\cal J}_1&=&\frac{1}{4} [ \omega_1(1-e_1)\left (
n_2(\omega_1)-n_1(\omega_1) \right )-(1\leftrightarrow 2)]
\end{eqnarray}
and ${\cal J}_2=-{\cal J}_1=-{\cal J}$. In LNEL, $\Delta T \ll 1$,
the Fourier's law is well verified, i.e. ${\cal J}=\kappa \Delta
T$, with $\kappa$ the thermal conductance depending on the qubit
internal splitting and mean temperature \cite{qrrp}. The evolution
of heat current and concurrence, as $\Delta T$ increases is
illustrated in Fig. 3 (obviously ${\cal J}=0$ for $\Delta T=0$).
Clearly, for temperatures for which $\alpha < 0$, see Fig. 2-a, an
enhancement of the concurrence is possible by applying a
temperature gradient. Based on this general concurrence's
behavior, we conclude that in LNEL, ${\cal J}(\Delta
T)=\frac{\kappa}{\sqrt{\alpha}}|C_{eq}-C(\Delta T)|^{1/2}$ as it
is clearly observed in Fig. 3. A remarkable point to be noted is
the possibility of constructing nonequilibrium quantum states with
identical concurrence, as that for the equilibrium case, but
carrying a heat current. It is also evident from Fig. 3 that the
relation between heat current and concurrence becomes independent
of $T_M$ as the temperature gradient increases. The correlation
between quantum linear entropy $S_L$ and concurrence is also shown
in the inset of Fig. 3, confirming the fact that a gradient
temperature will always increase the mixing degree of the quantum
state. Although, the amount of entanglement is small in those cases, it can
be significantly increased by distillation protocols.
\begin{figure}
\includegraphics[width=0.7\columnwidth]{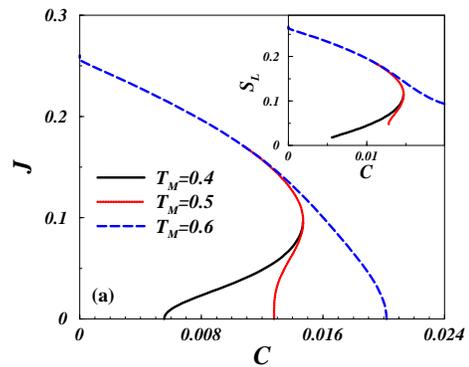}
\caption{ (Color online) Heat current ${\cal J}$ and concurrence
evolution for different mean temperatures $T_M$. Inset: Linear
entropy and concurrence. Each point corresponds to a $\Delta T$
value.}
\end{figure}
\begin{figure}
\includegraphics[width=0.7\columnwidth]{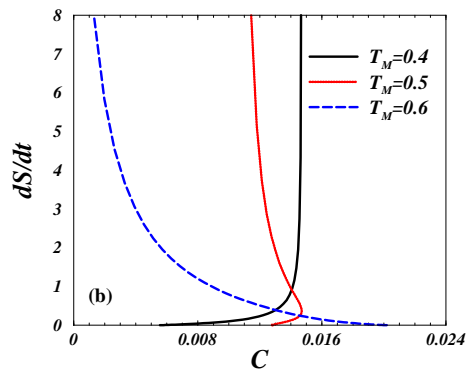}
\caption{ (Color online) Thermodynamical entropy production rate
and concurrence evolutions for different mean temperatures $T_M$.
Each point corresponds to a $\Delta T$ value.}
\end{figure}

Any heat current produces an amount of thermodynamical entropy
proportional to the heat which is carried on and inversely
proportional to the temperature of the reservoir from which the
heat is extracted (or injected). Thus, the thermodynamic entropy
production rate in our system can be written as $dS/dt={\cal
J}_1(T_1^{-1}-T_2^{-1})$ \cite{prigogine}. In LNEL, the rate of
entropy production shows a linear dependence with the concurrence
as $\frac{dS}{dt}=\frac{2\kappa}{\alpha T_M^2}|C_{eq}-C(\Delta
T)|$ as it is shown in Fig. 4. The entropy production rate, like
the heat current and linear entropy, is always different for two
different values of $\Delta T$ corresponding however to the same
amount of entanglement.
\begin{figure}
\includegraphics[width=0.7\columnwidth]{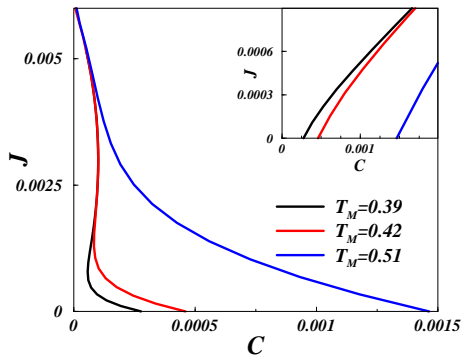}
\caption{ (Color online) Heat current ${\cal J}$ and concurrence
evolution for different mean temperatures $T_M$ in the
non-symmetric case, $\epsilon_1=8$ and $\epsilon_2=3$. Inset:
$\epsilon_1=3$ and $\epsilon_2=8$. Each point corresponds to a
$\Delta T$ value.}
\end{figure}

\subsection{B. Non-symmetric case,
$\epsilon_1>\epsilon_2$}
We first consider the high temperature
reservoir ($T_1$) is in direct contact with the large splitting
qubit, $\epsilon_1$, and the low temperature reservoir ($T_2$) is
in contact with the small splitting qubit, $\epsilon_2$.
Modifications to equilibrium values of physical magnitudes such as
the concurrence and the linear entropy are now of first order in
$\Delta T$ instead of order $\Delta T^2$, as it was the case for
the symmetric set up. This implies that in LNEL ${\cal J}(\Delta
T)\sim|C_{eq}-C(\Delta T)|$, as is illustrated in Fig. 5. Note
that at low temperature it arises the possibility of finding up to
three quantum states with the same concurrence but carrying
different heat currents. Switching to the inverse connection
between qubits and reservoirs ($T_1\leftrightarrow T_2$), the heat
current dependence on the concurrence is completely modified. In
this latter case, high temperature bath in contact with the low
splitting qubit, the heat current is substantially decreased but
the concurrence can be enhanced by the temperature gradient, as it
is shown in the inset of Fig. 5. We conclude that a qubit
splitting asymmetry brings an interesting new control parameter
for engineering nonequilibrium thermal quantum states.

\section{IV. Conclusions}
In summary, we have demonstrated that under nonequilibrium thermal
conditions a versatile scenario for tailoring heat carrying
quantum states with a well specified amount of entanglement is
feasible. A temperature gradient has been shown to produce
increasing or decreasing entanglement depending on the internal
coupling strength within a nanosystem.
Physical realizations of
the model system we addressed are provided by a large number of
physical systems such as nuclear spins in quantum dots and
superconducting qubits. Therefore, the resulting insights can
serve as useful recipes for realistic quantum information
processors in noisy and nonequilibrium environments.

L.Q. and F.J.R. acknowledge financial support from COLCIENCIAS and Facultad de
Ciencias-Uniandes-2006. M.E.R. and R. P. acknowledge financial support from
Vicerrector\'{\i}a Acad\'emica (2004-5), Universidad Javeriana-Bogot\'a.

\end{document}